\documentclass[
preprint,
amssymb,amsmath,aps,floatfix,prl,nofootinbib]{revtex4-1}
  \usepackage{latexsym,bm,amsmath,amssymb,amsfonts}
\newcommand{\beq}{\begin{eqnarray}}
\newcommand{\eeq}{\end{eqnarray}}

\newcommand{\nn}{\nonumber \\}

\usepackage{graphicx}
\usepackage{amsmath,amsthm,amssymb}

\begin{document}
\preprint{YITP-14-102}

\title{Use of the Husimi distribution for nucleon tomography}

\author{Yoshikazu Hagiwara}
\affiliation{Department of Physics, Kyoto University, Kyoto 606-8502, Japan}
\author{Yoshitaka Hatta}
\affiliation{Yukawa Institute for Theoretical Physics, Kyoto University, Kyoto 606-8502, Japan}

\date{\today}
\vspace{0.5in}
\begin{abstract}
In the context of nucleon structure, the Wigner distribution has been commonly used to visualize the phase-space distribution of quarks and gluons inside the nucleon. However, the Wigner distribution does not allow for a probabilistic interpretation because it takes negative values. In pursuit of a positive phase-space distribution in QCD, we introduce the Husimi distribution and demonstrate its advantages via a simple one-loop example. We also comment on a possible connection to the semiclassical approach to saturation physics at small-$x$.

\end{abstract}

\maketitle

\section{Introduction}

Multi-dimensional tomography has become an important paradigm in the modern study of the nucleon structure \cite{Boer:2011fh}.
Partons in a high energy nucleon are characterized not only by the longitudinal momentum fraction $x$, but also by the transverse momentum $\vec{k}_\perp$ and the transverse position $\vec{b}_\perp$. Such advanced knowledge is encoded in the transverse momentum dependent distribution (TMD) $T(x,\vec{k}_\perp)$ and (Fourier transform of) the generalized parton distribution (GPD) $G(x,\vec{b}_\perp)$. In addition to being indispensable for  calculating exclusive cross sections, these distributions nicely provide a visual way of understanding the  three-dimensional structure of the nucleon.

For certain purposes, however, information of \emph{both} $\vec{k}_\perp$ and $\vec{b}_\perp$ is needed. An example of great phenomenological interest is the orbital angular momentum $\vec{b}_\perp \times \vec{k}_\perp$ relevant to the nucleon spin decomposition. More generally, a fully-unintegrated distribution of the type $W(x,\vec{b}_\perp,\vec{k}_\perp)$ \cite{Ji:2003ak,Belitsky:2003nz,Lorce:2011kd} (see also \cite{Meissner:2009ww,Lorce:2013pza}) completely characterizes the nucleon wavefunction in terms of partons and serves as the `mother distribution' of TMDs and GPDs.
Such a joint distribution in the position and the momentum is well-known in quantum mechanics as the Wigner distribution \cite{Wigner}, and has long been used in a wide variety of contexts.  However, one should bear in mind that the very notion of `phase-space distribution' in quantum theory contradicts the uncertainty principle $\delta q \delta p\ge \hbar/2$. Because of this, the Wigner distribution is often violently oscillating and even becomes negative in some region of the phase space. The same problem is expected to persist in field theory. The QCD Wigner distribution as defined in \cite{Ji:2003ak,Belitsky:2003nz,Lorce:2011kd} is not positive-definite, and therefore it cannot be interpreted as the probability distribution of partons.

Fortunately, it is known that the Wigner distribution can be made positive-semidefinite by smearing it within the region of minimal uncertainty $\delta q \delta p= \hbar/2$. The \emph{Husimi distribution} \cite{husimi} thus obtained is the closest analog of the classical phase-space distribution that one can expect for a quantum system. In this paper we apply the idea of the Husimi distribution to QCD in pursuit of a positive phase-space distribution of partons.

We start by briefly reviewing the Husimi distribution in quantum mechanics. We then discuss the Wigner and Husimi distributions for QCD.  We shall demonstrate via a simple example how in practice the problem of negative regions in the Wigner distribution can be avoided in the Husimi distribution. Finally, we conclude with some speculations for future work.

\section{Husimi distribution in quantum mechanics}

Consider quantum mechanics in one dimension. For a generic pure quantum state $|\psi(t)\rangle$,
the Wigner distribution is defined as
\beq
f_W(q,p,t)&=&\int_{-\infty}^\infty dx e^{-ipx/\hbar} \langle \psi(t)|q-x/2\rangle \langle q+x/2|\psi(t)\rangle  \nn
&=&\int_{-\infty}^\infty dx e^{-ipx/\hbar}\langle q+x/2|\hat{\rho}(t)|q-x/2\rangle\,, \label{1}
\eeq
 where $\hat{\rho}(t)=|\psi(t)\rangle \langle \psi(t)|$ is the density matrix. $f_W$ is a function of both the position $q$ and the momentum $p$, and satisfies the following conditions
 \beq
\int \frac{dq}{2\pi\hbar}f_W(q,p,t)=|\langle \psi(t)|p\rangle|^2\,,&&\quad \int \frac{dp}{2\pi\hbar}f_W(q,p,t)=|\langle \psi(t)|q\rangle|^2\,, \label{pro} \\ \int \frac{dqdp}{2\pi \hbar} &&f_W(q,p,t)=1\,.
 \eeq
 From (\ref{pro}), it is tempting to interpret $f_W$ as the probability distribution in the phase space $(q,p)$. However,
 it cannot be literally interpreted as such, since it is not positive-definite and often violently oscillating. The reason of this failure is the uncertainty principle  which nullifies any attempt to simultaneously measure the position and momentum beyond the accuracy $\delta q\delta p\ge \hbar/2$. In quantum mechanics, the best one can do is to speak of the probability of finding a particle within the band $(q\pm \delta q/2, p\pm \delta p/2)$ of minimal uncertainty $\delta q \delta p=\hbar/2$. This is achieved by the Husimi distribution \cite{husimi} which is the Gaussian convolution of the Wigner distribution
\beq
f_H(q,p,t)=\frac{1}{\pi\hbar}\int dq'dp'e^{-m\omega(q'-q)^2/\hbar - (p'-p)^2/m\omega\hbar} f_W(q',p',t)\,, \label{hu}
\eeq
 where $m$ is the particle mass and $\omega$ is an arbitrary parameter.
 The widths of the Gaussian factors indicate that the distribution is smeared in the position space $\delta q=\sqrt{\hbar/2m\omega}$ and  the momentum space $\delta p=\sqrt{\hbar m\omega/2}$ such that $\delta q \delta p=\hbar/2$. Different values  of $\omega$ correspond to different resolution scales with which one probes the system. For oscillator systems including radiation fields, $\omega$ is often identified with the frequency \cite{lee}.\footnote{For the harmonic oscillator, the Wigner and Husimi distributions can be computed analytically. For the $n$-th excited state, the results are
  \beq
  f_W^{(n)}(q,p)=2(-1)^n e^{-\frac{2H}{\hbar \omega}}L_n\left(\frac{4H}{\hbar\omega}\right)\,, \qquad f_H^{(n)}(q,p)=\frac{1}{n!}e^{-\frac{H}{\hbar\omega}}\left(\frac{H}{\hbar\omega}\right)^n\,,
 \label{harm} \eeq
  where $L_n$ is the Laguerre polynomial and $H=\frac{p^2}{2m}+\frac{m\omega^2q^2}{2}$ is the classical Hamiltonian. The parameter $\omega$ in (\ref{hu}) is identified with the frequency of the harmonic oscillator. In contrast to the Wigner distribution which has unphysical oscillations, the Husimi distribution is manifestly positive-semidefinite and localized near  the classical orbit $H\approx \hbar\omega n$.}

A remarkable property of the Husimi distribution is that
 it is positive-semidefinite
 \beq
 f_H(q,p,t)=\langle \lambda|\hat{\rho}|\lambda\rangle=|\langle \psi|\lambda\rangle|^2\ge 0\,,
 \label{posi}
 \eeq
 where $|\lambda\rangle=e^{\lambda\hat{a}^\dagger-\lambda^*\hat{a}}|0\rangle$  is the coherent state which is the eigenstate of the annihilation operator $\hat{a}|\lambda\rangle =\lambda|\lambda\rangle$. [We defined $\lambda=\frac{m\omega q+ip}{\sqrt{2\hbar m\omega}}$ and $\hat{a}=\frac{1}{\sqrt{2\hbar m\omega}}(m\omega \hat{q}+i\hat{p})$.]  The coherent state is often referred to as the `most classical' quantum state since it realizes the minimal uncertainty relation
$\delta q\delta p=\hbar/2$ for any value of $\omega$.\footnote{The original state $|\psi\rangle$ generically has a larger uncertainty. Taking the expectation value in the coherent state (\ref{posi}) simply means that  we look at the system with a coarse-grained filter with $\delta q\delta p=\hbar/2$.  } It is then natural that the Gaussian convolution (\ref{hu}) is mathematically equivalent to introducing the coherent state.

 Thanks to the positivity and the normalization condition $\int \frac{dqdp}{2\pi \hbar} f_H(q,p,t)=1$, the Husimi distribution can be legitimately  interpreted as a probability distribution.
  It has numerous applications in statistical physics, condensed matter physics, quantum optics, quantum chaos, and also in atomic and nuclear physics \cite{lee,Kunihiro:2008gv}.

\section{Husimi distribution in QCD}

We now turn to QCD. In the context of nucleon structure, the Wigner distribution for quarks is defined by (setting $\hbar=c=1$) \cite{Belitsky:2003nz,Lorce:2011kd}
\beq
W(x, \vec{b}_\perp, \vec{k}_\perp) &=&\int \frac{dz^- d^2z_\perp}{16\pi^3} \frac{d^2\Delta_\perp}{(2\pi)^2}e^{i(xP^+z^- -\vec{k}_\perp \cdot \vec{z}_\perp)}e^{-i\vec{\Delta}_\perp \cdot \vec{b}_\perp} \nonumber \\
  && \qquad  \times\langle P+\Delta/2| \bar{\psi}(-z/2)\Gamma {\mathcal L} \psi(z/2)|P-\Delta/2\rangle \nn
&=& \int \frac{dz^- d^2z_\perp}{16\pi^3} \frac{d^2\Delta_\perp}{(2\pi)^2}e^{i(xP^+z^- -\vec{k}_\perp \cdot \vec{z}_\perp)} \nonumber \\ && \qquad \times \langle P+\Delta/2| \bar{\psi}(b -z/2)\Gamma {\mathcal L}\psi(b +z/2)|P-\Delta/2\rangle \,, \label{wigner}
\eeq
 where $P^\mu$ is the nucleon momentum and $z^\mu=(0,z^-,\vec{z}_\perp)$, $\Delta^\mu=(0,0,\vec{\Delta}_\perp)$, $b^\mu=(0,0,\vec{b}_\perp)$.  ${\mathcal L}$ is the staple-shaped Wilson line along the light-cone $z^-$ that makes the operator gauge invariant. $\Gamma$ is some gamma matrix $\gamma^+,\gamma^+\gamma_5$, etc. In what follows, for definiteness we only consider the case $\Gamma=\gamma^+$ relevant to the unpolarized distribution. One can also define the Wigner distribution for gluons in a similar manner.

 Eq.~(\ref{wigner}) describes the transverse phase-space distribution in the position $\vec{b}_\perp$ and the momentum $\vec{k}_\perp$ of quarks carrying the longitudinal momentum fraction $x$.  It is the `mother function' of well-known distributions in QCD:
Integrating over $\vec{b}_\perp$, one gets the transverse momentum distribution (TMD)
\beq
\int d^2b_\perp W(x, \vec{b}_\perp, \vec{k}_\perp) &=& \int \frac{dz^- d^2z_\perp}{16\pi^3} e^{i(xp^+z^- -\vec{k}_\perp \cdot \vec{z}_\perp)}\langle P| \bar{\psi}(-z/2)\gamma^+ {\mathcal L} \psi(z/2)|P\rangle \,. \label{tmd}
\eeq
Integrating over $\vec{k}_\perp$, one gets the Fourier transform of the generalized parton distribution (GPD)
\beq
&&   \int d^2k_\perp W(x, \vec{b}_\perp, \vec{k}_\perp) \nn && =\int \frac{d^2\Delta_\perp}{(2\pi)^2}e^{-i\vec{\Delta}_\perp \cdot \vec{b}_\perp}\int \frac{dz^-}{4\pi} e^{ixp^+z^-} \langle P+\Delta/2| \bar{\psi}(-z^-/2)\gamma^+ {\mathcal L} \psi(z^-/2)|P-\Delta/2\rangle\,.
\eeq
For the longitudinally polarized nucleon, the Wigner distribution is also related to the \emph{canonical} orbital angular momentum \cite{Lorce:2011kd,Hatta:2011ku}
\beq
L_W= \int dx d^2b_\perp d^2k_\perp (\vec{b}_\perp\times \vec{k}_\perp) W (x, \vec{b}_\perp, \vec{k}_\perp)\,, \label{oam}
\eeq
 which is an important ingredient in the nucleon spin decomposition.

The Wigner distribution (\ref{wigner}) has been evaluated in various models \cite{Belitsky:2003nz,Lorce:2011kd,Lorce:2011ni,Kanazawa:2014nha,Mukherjee:2014nya,
Liu:2014vwa,Muller:2014tqa,Courtoy:2014bea}. In simple models without gluons, it turns out to be a positive function. However, once gluons are included, it becomes negative in some region of the phase space \cite{Mukherjee:2014nya}.
This motivates us to introduce the QCD version of the Husimi distribution. Similarly to (\ref{hu}), we try
\beq
H(x, \vec{b}_\perp, \vec{k}_\perp)&\equiv & \frac{1}{\pi^2}\int d^2b'_\perp d^2k'_\perp e^{-\frac{1}{\ell^2}(\vec{b}_\perp-\vec{b}'_\perp)^2 -\ell^2(\vec{k}_\perp-\vec{k}'_\perp)^2}W(x, \vec{b}'_\perp, \vec{k}'_\perp) \nn
&=&  \int \frac{dz^- d^2z_\perp}{16\pi^3} \frac{d^2\Delta_\perp}{(2\pi)^2}e^{i(xp^+z^- -\vec{k}_\perp \cdot \vec{z}_\perp)}e^{-i\vec{\Delta}_\perp \cdot \vec{b}_\perp} e^{-\ell^2\frac{\Delta_\perp^2}{4}-\frac{z_\perp^2}{4\ell^2}} \nn
&& \qquad \times \langle P+\Delta/2| \bar{\psi}(-z/2)\gamma^+ {\mathcal L}\psi(z/2)|P-\Delta/2\rangle  \,, \label{hu2}
\eeq
 where the length parameter $\ell$ is in principle arbitrary, but it is natural to choose $\ell \lesssim R_h$ with $R_h$ being the hadronic radius (except in the small-$x$ region, see a later discussion). Eq.~(\ref{hu2}) seems to be a reasonable extension of (\ref{hu}) to the field theory case, but it is \emph{a priori} not obvious whether the above definition gives a positive-definite distribution. The main concern is that, unlike in nonrelativistic quantum mechanics, the initial and final states are different due to the momentum recoil $\Delta_\perp \neq 0$ necessary to probe the nucleon \cite{Belitsky:2003nz}. Unfortunately, this is inevitable in a relativistic system, and as a result the normal proof of positivity (\ref{posi}) cannot be used. Nevertheless,  there is a good chance that (\ref{hu2}) is positive in the physically interesting region as we argue now.\footnote{To avoid misunderstanding, we note that here we are talking about the positivity of the phase space distribution of the quark number. The electromagnetic charge distribution can of course be negative if the quark has negative charge. }

Let us work in the infinite momentum frame $p^+\gg 1/R_h$ and in the light-cone gauge $A^+=0$ where the Wilson line can be set to unity.  Then we can formally write
 \beq
 W &\sim&\! \int \!d^2\Delta_\perp\, e^{-i\vec{\Delta}_\perp \cdot \vec{b}_\perp}\langle P\!+\!\Delta/2|\psi_+^\dagger\delta(K^+\!-\!(1\!-\!x)p^+)\delta^{(2)}
 (\vec{K}_\perp+\vec{k}_\perp)\psi_+|P\!-\!\Delta/2\rangle \,, \\
 H &\sim& \!\int\!d^2\Delta_\perp\, e^{-i\vec{\Delta}_\perp \cdot \vec{b}_\perp-\frac{\ell^2 \Delta_\perp^2}{4}}\langle P\!+\!\Delta/2|\psi_+^\dagger\delta(K^+\!-\!(1\!-\!x)p^+) e^{-
 \ell^2(\vec{K}_\perp+\vec{k}_\perp)^2}\psi_+|P\!-\!\Delta/2\rangle\,,\nonumber
 \eeq
 where $K^\mu$ is the momentum (translation) operator and $\psi_+\equiv \frac{1}{2}\gamma^-\gamma^+\psi(0)$ is the so-called `good component' of the quark field \cite{Jaffe:1996zw}. We see that the operator involved is positive-definite in both cases, but the Fourier transform of its \emph{off}-diagonal matrix element does not have a definite sign. Nevertheless, we clearly see qualitative differences in the two cases. In particular, unlike in the Wigner distribution, the $d^2\Delta_\perp$ integral of the Husimi distribution is bounded $|\Delta_\perp|/2 \lesssim 1/\ell\sim 1/R_h$ because of the Gaussian damping. This makes the matrix elements nearly diagonal and therefore the integrand of the Fourier transform is dominantly positive, resulting in a positive distribution.  It should be said that we cannot  rigorously exclude the possibility of finding negative regions in corners of the phase space. However, our argument strongly disfavors such a possibility. In the next section, we shall demonstrate how in practice the negative Wigner distribution is converted into the  positive Husimi distribution using an explicit model.

In quantum mechanics, it is known that the $q$- or $p$-moment of the Husimi distribution $f_H(q,p)$ does not reduce to the probability distribution (\ref{pro}) in a single variable $p$ or $q$. The same thing happens here.
The $\int d^2b_\perp$ or $\int d^2k_\perp$ moment of  $H$ does not reduce to a known distribution. For example,
\beq
\int d^2b_\perp H(x, \vec{b}_\perp, \vec{k}_\perp)&=&
  \int \frac{dz^- d^2z_\perp}{16\pi^3} e^{i(xp^+z^- -\vec{k}_\perp \cdot \vec{z}_\perp)} e^{-\frac{z_\perp^2}{4\ell^2}}  \langle P| \bar{\psi}(-z/2)\gamma^+ {\mathcal L}\psi(z/2)|P\rangle\,. \label{tt}
\eeq
This is similar to the TMD (\ref{tmd}), but a Gaussian regularization factor is inserted. We shall later comment on the possible interpretation of this factor.   On the other hand, the double moment $\int d^2b_\perp d^2k_\perp$ gives the ordinary parton distribution as in the case of the Wigner distribution
\beq
\int \!d^2b_\perp d^2k_\perp H = \int \!d^2b_\perp d^2k_\perp W=\int \frac{dz^-}{4\pi}e^{ixP^+z^-}\langle P|\bar{\psi}(-z^-/2)\gamma^+ {\mathcal L}\psi(z^-/2)|P\rangle\,.
\eeq
Similarly, for the canonical orbital angular momentum it is easy to see that
\beq
L_H &\equiv& \int dx d^2b_\perp d^2k_\perp (\vec{b}_\perp\times \vec{k}_\perp) H (x, \vec{b}_\perp, \vec{k}_\perp) \\
&=& - \int  \frac{dx dz^-}{4\pi}e^{ixp^+z^-}  \left(\frac{\partial}{\partial \vec{\Delta}_\perp}\times \frac{\partial}{\partial \vec{z}_\perp}\right) \nn
&& \qquad \times\left.  e^{-\frac{\ell^2\Delta_\perp^2}{4}-\frac{z_\perp^2}{4\ell^2}} \langle P+\Delta/2| \bar{\psi}(-z/2)\gamma^+{\mathcal L} \psi(z/2)|P-\Delta/2\rangle\right|_{\Delta_\perp=z_\perp=0}\,. \nonumber
\eeq
Clearly, the Gaussian factors are irrelevant so that $L_H=L_W$. Since the Husimi distribution is much better-behaved than the Wigner distribution (see the next section), the computation of $L$ in lattice QCD \cite{Ji:2013dva} via the Husimi distribution may be numerically more stable.

\section{One-loop example}

 As an illustration, let us compute the Husimi distribution for a single quark dressed by a gluon at one-loop order. This example is simple enough so that the corresponding Wigner distribution can be calculated analytically. Yet it illuminates the nontrivial issue of how the positivity, violated in the Wigner distribution, is restored in the Husimi distribution.

  Consider an unpolarized, on-shell quark with mass $m$. To zeroth order, the Wigner and Husimi distributions have support only at $x=1$
 \beq
 W(x,\vec{b}_\perp,\vec{k}_\perp)&=&\delta(x-1)\delta^{(2)}(\vec{b}_\perp)
\delta^{(2)}(\vec{k}_\perp)\,, \label{www}  \\
 \quad \Longrightarrow H(x,\vec{b}_\perp,\vec{k}_\perp)&=&
\delta(x-1)  \frac{e^{-b_\perp^2/\ell^2-\ell^2k_\perp^2}}{\pi^2}\,.
\eeq
Note that the product of the delta functions in the Wigner distribution (\ref{www}) forces $\vec{b}_\perp=\vec{k}_\perp=0$, and this implies a violation of the uncertainty principle mentioned in the introduction.\footnote{$\vec{b}_\perp$ and $\vec{k}_\perp$ are subject to the uncertainty principle although they are not Fourier-conjugate variables \cite{Lorce:2011kd}. Actually, the situation is the same in quantum mechanics. In (\ref{1}), $p$ and $x$ are Fourier-conjugate, but one speaks of  uncertainty  in $p$ and $q$. In the present example, one can trivially obtain  the (genuine) probability distributions of $\vec{k}_\perp$ and $\vec{b}_\perp$ from the Wigner distribution, $\int db_\perp W(b_\perp,k_\perp)\sim \delta(k_\perp)$ and  $\int dk_\perp W(b_\perp,k_\perp)\sim \delta(b_\perp)$. This unambiguously shows  the violation of the uncertainty principle.  }
This has been remedied in the Husimi distribution.

At one-loop order, the Wigner distribution is most conveniently calculated in the light-cone gauge $A^+=0$. For simplicity, we assume $x<1$ in the following. The result is  \cite{Mukherjee:2014nya} (see also \cite{Miller:2014vla})
\beq
W(x,\vec{b}_\perp,\vec{k}_\perp)=\frac{\alpha_s C_F}{2\pi^2} \int \frac{d^2\Delta_\perp}{(2\pi)^2}
e^{-i\vec{\Delta}_\perp \cdot \vec{b}_\perp}
\frac{\vec{q}_+\cdot \vec{q}_- P_{qq}(x)+m^2(1-x)^3}{(q_+^2+m^2(1-x)^2)(q_-^2+m^2(1-x)^2)}\,, \label{w}
\eeq
 where  $P_{qq}(x)=\frac{1+x^2}{1-x}$ is the splitting function and we defined $\vec{q}_\pm \equiv \vec{k}_\perp \pm \frac{\vec{\Delta}_\perp}{2}(1-x)$.
One immediately recognizes some undesirable features in (\ref{w}). Firstly, the $d^2\Delta_\perp$ integral is logarithmically divergent for $\vec{b}_\perp=0$ and converges very slowly for $|\vec{b}_\perp|\neq 0$. In practice, a cutoff is needed at $|\vec{\Delta}_\perp|=\Delta_\perp^{max}$ and the result depends on $\Delta_\perp^{max}$ rather strongly \cite{Mukherjee:2014nya}. Secondly, the coefficient of $P_{qq}$ turns negative when
\beq
|\vec{k}_\perp|< (1-x) \frac{|\vec{\Delta}_\perp|}{2}\sim \frac{1-x}{2|\vec{b}_\perp|}\,,
\label{da}
\eeq
 and in this regime the Wigner distribution indeed becomes negative  unless $m$ is large or $x\approx 1$. Thirdly, the factor $e^{-i\vec{\Delta}_\perp \cdot \vec{b}_\perp}$ oscillates rapidly at large $|\vec{b}_\perp |$, providing another source of the negative values of the Wigner distribution. These features  reflect the quantum interference effect encoded in the Wigner distribution. However, it is  counter-intuitive to find negative regions of the quark number distribution in this one-loop model where there is no antiquark. Moreover, the probabilistic interpretation is not possible.

 We now argue that all of these problems can be resolved by switching to the Husimi distribution
\beq
H(x,\vec{b}_\perp,\vec{k}_\perp)&=&
 \ell^2\frac{\alpha_s C_F}{2\pi^3} \int d^2k'_\perp e^{-\ell^2(\vec{k}_\perp-\vec{k}'_\perp)^2} \int \frac{d^2\Delta_\perp}{(2\pi)^2} \cos(\vec{\Delta}_\perp \cdot \vec{b}_\perp)e^{-\frac{\ell^2}{4}\Delta_\perp^2} \nn && \qquad \qquad \times\frac{\vec{q_+}'\cdot \vec{q_-}' P_{qq}(x)+m^2(1-x)^3}{((q'_+)^2+m^2(1-x)^2)((q'_-)^2+m^2(1-x)^2)}\,, \label{h}
\eeq
 where we have taken the real part knowing that the Wigner and hence Husimi distributions are real \cite{Lorce:2011kd}.
The $d^2\Delta_\perp$ integral is effectively cut off at $|\vec{\Delta}_\perp|\lesssim 2/\ell$ so that there is no convergency problem.
At the same time, smearing within the region $|\vec{k}_\perp-\vec{k}'_\perp| \lesssim \frac{1}{\ell}$ completely encompasses the dangerous region (\ref{da})
\beq (1-x)\frac{|\vec{\Delta}_\perp|}{2} \le \frac{|\vec{\Delta}_\perp|}{2} \lesssim \frac{1}{\ell} \,.
\eeq
This ensures that the negative contribution from (\ref{da}) is canceled by the positive contribution from the surrounding region in much the same way as in quantum mechanics.

On the other hand, the third problem is of relativistic origin and not present in nonrelativistic quantum mechanics as we already warned below (\ref{hu2}). Still, the Husimi distribution can handle this.
  The oscillating factor $\cos(\vec{\Delta}_\perp \cdot \vec{b}_\perp)$ first turns negative when  $|\vec{\Delta}_\perp \cdot \vec{b}_\perp|>\frac{\pi}{2}$, and the successive negative regions have the size $|\delta b_\perp|=\frac{\pi}{|\vec{\Delta}_\perp|}<2|\vec{b}_\perp|$.
The smearing in the $\vec{b}_\perp$-space is performed in the region $|\delta b_\perp|<2\ell$, so when $|\vec{b}_\perp|<\ell$, again there will be a cancellation. On the other hand, when $|\vec{b}_\perp|\gg \ell$,  $H$ is exponentially suppressed as $e^{-b_\perp^2/\ell^2}$ (see (\ref{hu2})).\footnote{This being said, we cannot exclude the possibility that the relativistic Husimi distribution slightly becomes negative in the region $|\vec{b}_\perp| \gtrsim \ell$. When this happens, we can choose a large enough value of $\ell$ such that the negative region is relegated far away from the quark.}

 \begin{figure}[htbp]
\begin{minipage}{0.45\hsize}
\begin{flushleft}
   \includegraphics[width=90mm]{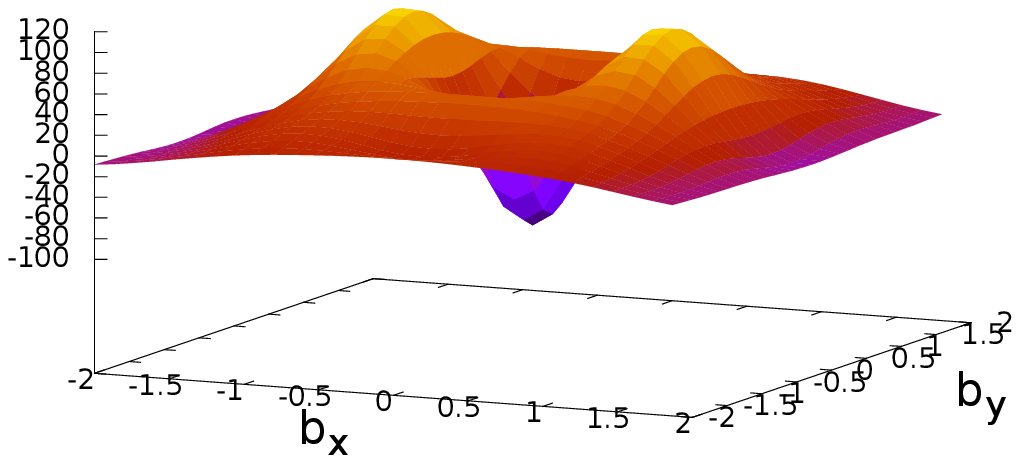}
 \end{flushleft}
 \end{minipage}
 \begin{minipage}{0.45\hsize}
   \includegraphics[width=90mm]{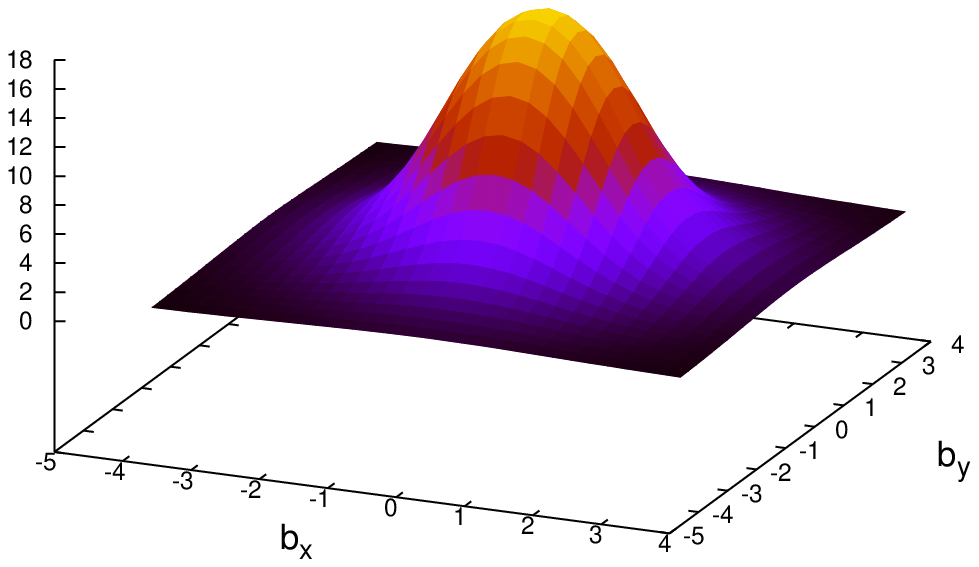}
 \end{minipage}
 \caption{Plots of the Wigner (left) and Husimi (right) distributions in the $\vec{b}_\perp$-space at $x=0.5$. Here and in Fig.~\ref{fig2}, the units of the horizontal axes are in GeV$^{-1}$.\label{fig1}}
\end{figure}
  \begin{figure}[htbp]
\begin{minipage}{0.45\hsize}
\begin{flushleft}
   \includegraphics[width=90mm]{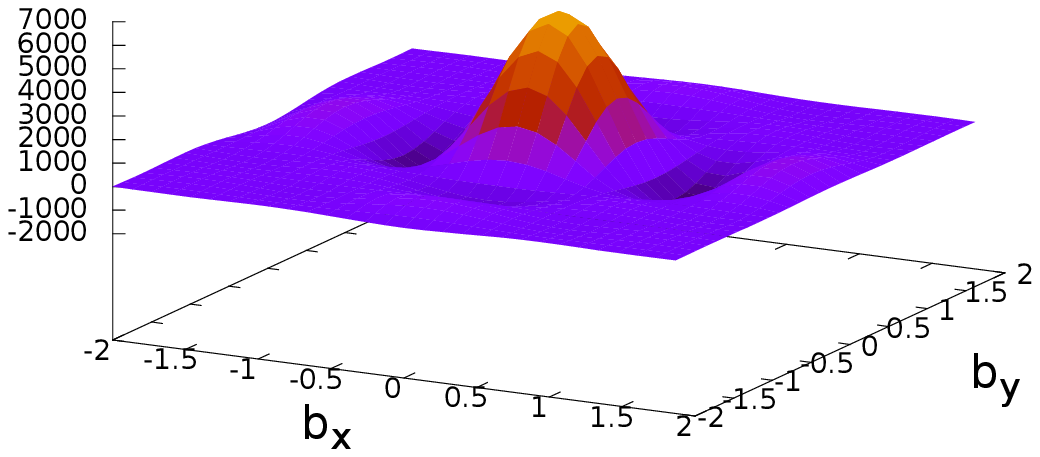}
 \end{flushleft}
 \end{minipage}
 \begin{minipage}{0.45\hsize}
   \includegraphics[width=90mm]{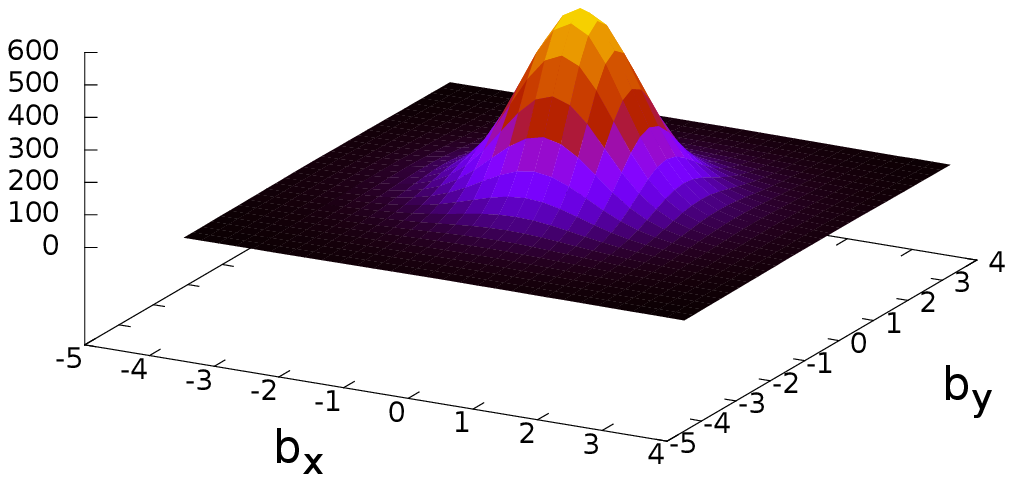}
 \end{minipage}
 \caption{Plots of the Wigner (left) and Husimi (right) distributions in the $\vec{b}_\perp$-space at $x=0.9$. \label{fig2}}
\end{figure}

   \begin{figure}[htbp]
\begin{minipage}{0.45\hsize}
\begin{flushleft}
   \includegraphics[width=90mm]{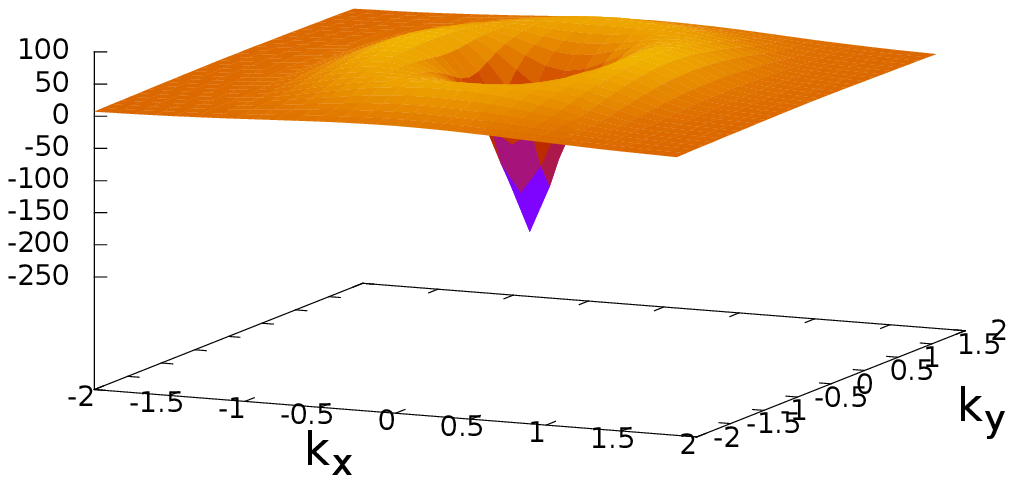}
 \end{flushleft}
 \end{minipage}
 \begin{minipage}{0.45\hsize}
   \includegraphics[width=90mm]{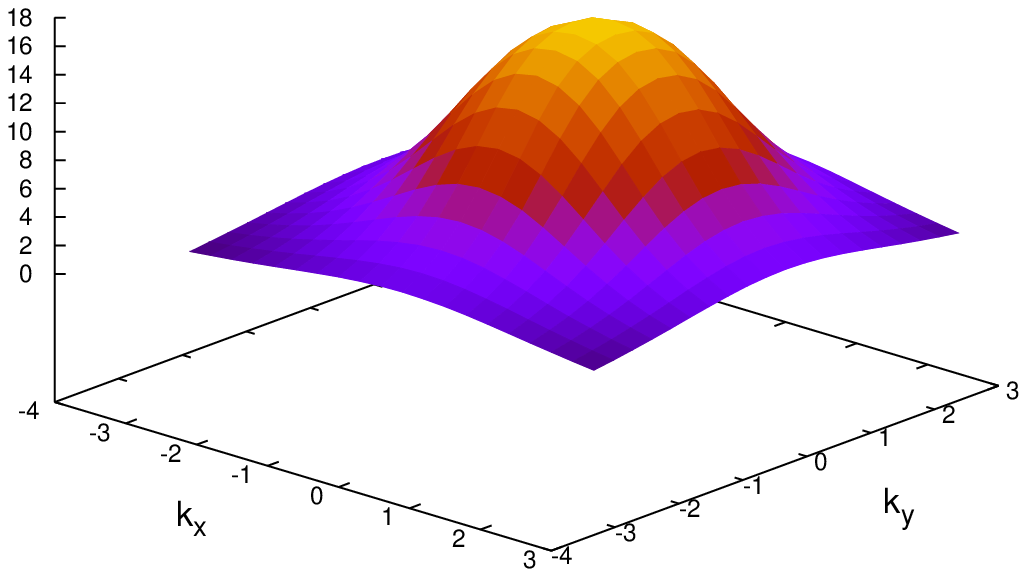}
 \end{minipage}
 \caption{Plots of the Wigner (left) and Husimi (right) distributions in the $\vec{k}_\perp$-space at $x=0.5$. The units of the horizontal axes are in GeV. \label{fig3}}
\end{figure}
In order to make these arguments quantitative, we must resort to numerical methods.
In Fig.~\ref{fig1}, we show the Wigner and Husimi distributions (divided by the common prefactor $\frac{\alpha_s C_F}{2\pi^2(2\pi)^2}$) in the $\vec{b}_\perp$-space at fixed $\vec{k}_\perp=(0.5\mbox{\,GeV},0)$.  We choose the parameters $x=0.5$, $m^2=0.1\, \mbox{GeV}^2$, $\ell = 1\,\mbox{GeV}^{-1}$ and $\Delta_\perp^{max}=5\,\mbox{GeV}$.
Fig.~\ref{fig2} is the same as Fig.~\ref{fig1} except that  $x=0.9$.
In Fig.~\ref{fig3} we show the two distributions at $x=0.5$ in the $\vec{k}_\perp$-space at fixed $\vec{b}_\perp=(0.5\mbox{\,GeV}^{-1},0)$.
Clearly, the Wigner distribution is nowhere near what one would expect for a phase-space distribution. Wiggles and negative peaks  are actually quite common in the Wigner distribution and often have no physical meaning. (One sees such unphysical behaviors already in the harmonic oscillator case (\ref{harm}).) In contrast, the Husimi distribution is well-behaved and turns out to be always positive as we have expected. (We tested other sets of parameters and did not find any negative regions as far as we could see.)  Therefore, at least in our chosen model the Husimi distribution can be interpreted as the phase-space \emph{probability} distribution of quarks at a given value of $x$.

\section{Possible connection to saturation physics}

For the single quark problem, there is not a natural value of $\ell$ to be used in the Husimi distribution (\ref{h}). It is just a free parameter associated with our choice of the resolution scale $\delta b_\perp \sim \ell$, $\delta k_\perp \sim 1/\ell$ to probe the system.  In the case of the nucleon, our preferred choice  is the nucleon radius  $\ell\lesssim R_h$ as we remarked already. On the other hand, for the gluon distribution at small-$x$, a very natural choice would be $\ell =1/Q_s(x)$ where $Q_s(x)$ is the saturation scale which becomes perturbative at small-$x$ and/or for a large nucleus \cite{McLerran:1993ni,yuri}. Indeed, $1/Q_s$ is the length scale beyond which the gluons can be treated coherently as a classical field. With this choice, it is interesting to notice that the factor  $e^{-z_\perp^2/4\ell^2}$ which accompanies the unintegrated gluon distribution (consider the gluonic version of (\ref{tt}), $\bar{\psi}\psi \to F^{+\mu}F^+_{\ \mu}$) becomes formally identical to the so-called forward dipole amplitude $e^{-Q_s^2z_\perp^2/4}$ often encountered in the semiclassical evaluation of the nucleon (or nucleus) matrix element. We thus conjecture that the notion of the Husimi distribution as the coherent state expectation value of the density matrix is smoothly connected to the semiclassical approach to saturation physics at small-$x$. In other words, what is calculated via classical gluon fields  could be reinterpreted as the Husimi distribution.

\section{Conclusion}

In this paper we have proposed the use of the Husimi distribution for nucleon tomography as an alternative to the often badly-behaved Wigner distribution. To support this idea, we used a simple one-loop model and demonstrated, both by argument and numerically, that the Wigner distribution which takes negative values can indeed be made  positive by transforming to the Husimi distribution. While the positivity of the Husimi distribution is well-known in statistical physics, a demonstration of this in the context of  relativistic field theory is nontrivial and new.
 In future, it is important to use more realistic models of the nucleon with multiple gluons and extend to the gluon distribution function. Including various polarization effects as in \cite{Lorce:2011kd,Mukherjee:2014nya,Liu:2014vwa} is also interesting. Finally we speculated about  a possible connection to saturation physics which may deserve further investigations.\\

\noindent
\textbf{Acknowledgements}

\noindent
We thank Ayumu Sugita and Teiji Kunihiro for an introduction to the Husimi distribution. We also thank Barbara Pasquini for discussion and Sreeraj Nair for correspondence. Numerical computation in this work was carried out at the Yukawa Institute Computer Facility.

\end{document}